\documentclass[conference]{IEEEtran}
\usepackage{graphicx} 
\usepackage{indentfirst}
\usepackage{algpseudocode}
\usepackage{amsmath}
\usepackage{amsfonts}
\usepackage{amssymb}
\usepackage{multirow}
\usepackage{algorithm}
\usepackage[font=small]{caption}
\usepackage{cite}
\usepackage{pifont}
\usepackage{makecell}
\usepackage{subcaption}
\usepackage{enumitem}
\usepackage{float}
\usepackage{url}
\def\BibTeX{{\rm B\kern-.05em{\sc i\kern-.025em b}\kern-.08em T\kern-.1667em\lower.7ex\hbox{E}\kern-.125emX}}
\title{Toward Agentic AI: Task-Oriented Communication for Hierarchical Planning of Long-Horizon Tasks}
\author{
    \IEEEauthorblockN{Sin-Yu Huang, Lele Wang, and Vincent W.S. Wong}
        Department of Electrical and Computer Engineering, The University of British Columbia, Vancouver, Canada\\
        Email: \{syhuang, lelewang, vincentw\}@ece.ubc.ca
}

\usepackage{xcolor}

\begin{document}

\maketitle

\begin{abstract} 
Agentic artificial intelligence (AI) is an AI paradigm that can perceive the environment, reason over observations, and execute actions to achieve specific goals. Task-oriented communication supports agentic AI by transmitting only the task-related information instead of full raw data in order to reduce the bandwidth requirement. In real-world scenarios, AI agents often need to perform a sequence of actions to complete complex tasks. Completing these long-horizon tasks requires a hierarchical agentic AI architecture, where a high-level planner module decomposes a task into subtasks, and a low-level actor module executes each subtask sequentially. Since each subtask has a distinct goal, the existing task-oriented communication schemes are not designed to handle different goals for different subtasks. To address this challenge, in this paper, we develop a hierarchical task-oriented communication (HiTOC) framework. We consider a system with an edge server and a robot as an edge device. The high-level planner and low-level actor modules reside on the edge server. The robot transmits only the environmental observation that is relevant to the current subtask to the edge server. We propose a conditional variational information bottleneck (cVIB) approach to train the HiTOC framework to adaptively transmit minimal information required for each subtask. Simulations conducted on the AI2-THOR platform demonstrate that the proposed HiTOC framework outperforms three state-of-the-art schemes in terms of the success rate on MAP-THOR benchmark.
\end{abstract}

\section{Introduction}
\label{Sec:intro}
Agentic artificial intelligence (AI) refers to AI systems that possess the capability to perceive the environment, reason over observations, and execute actions to achieve a specific goal~\cite{Durante2024,Jiang2025}. For example, autonomous driving relies on an agentic AI system to perceive the surroundings, reason about road conditions, and generate actions for driving. 
In practical systems, high-level tasks such as navigating to a destination can be decomposed into subtasks of route planning, lane changing, and passing intersections. Each subtask itself requires a sequence of actions. These long-horizon tasks are often too complex for a single large language model (LLM) to choose all the sequential actions~\cite{Durante2024}. Thus, recent studies have proposed hierarchical agentic AI architecture, in which a high-level planner module decomposes a task into multiple subtasks. A low-level actor module completes each subtask sequentially~\cite{Durante2024,Nayak2024}. For example, in~\cite{Nayak2024}, the LLaMAR architecture employs a plan-act-correct-verify strategy that can decompose a task into subtasks, select the actions, and conduct self-correction based on feedback. Although the existing AI agents demonstrate strong capabilities to complete long-horizon tasks, they rely on continuous updates of the environment to make decisions. 
In large-scale systems where an AI agent coordinates with numerous distributed entities, environment updates over wireless networks can lead to significant bandwidth usage.




Task-oriented communication addresses the aforementioned challenge by only transmitting information that is relevant to the task rather than sending the entire raw data. This approach can reduce bandwidth usage without performance degradation of the task~\cite{Diao2025,Li2025,Huang2025-2}. In~\cite{Diao2025}, the aligned task and reconstruction-oriented communication (ATROC) framework was proposed to transmit minimal information required for an AI agent to make correct decisions. However, in hierarchical agentic AI systems, different subtasks rely on distinct information. The existing task-oriented communication schemes are mainly developed for tasks with a single goal. The question of how to adaptively transmit the relevant information for each subtask remains an open challenge. Even in recent studies that considered multi-task scenarios~\cite{Huang2025}, how to generate minimal task-specific information to specify the intended task remains unexplored.

The aforementioned research gap motivates our work. In this paper, we propose a task-oriented communication framework that integrates a conditional module to extract minimal subtask information and a joint source–channel coding (JSCC) module to adaptively transmit minimal information required to complete each subtask. The main contributions are as follows:
\begin{itemize}
    \item We design a hierarchical task-oriented communication (HiTOC) framework. Upon receiving a task instruction, the high-level planner LLM decomposes the task into a sequence of subtasks. For each subtask, a conditional module on the edge server extracts the subtask information and sends it to the robot. The robot uses a JSCC encoder to encode its observation conditioned on the received subtask information. The encoded representation is then transmitted to the edge server and decoded by the JSCC decoder, which is also conditioned on the subtask information. By conditioning the encoding and decoding processes on the subtask information, the proposed HiTOC framework adaptively focuses on the information which is relevant to each subtask.
    \item We further propose a conditional variational information bottleneck (cVIB) approach to jointly train the conditional module, JSCC encoder, and JSCC decoder within the HiTOC framework. The goal is to train a conditional module that extracts the minimal subtask information to define the current goal. The JSCC encoder and decoder are trained to encode the input into a latent representation and reconstruct the task-specific image, respectively, both conditioned on the subtask. This enables the low-level actor LLM to make correct decisions. We define a loss function based on the conditional information bottleneck to jointly achieve these objectives and derive the variational upper bound.
    \item We conduct simulations on AI2-THOR~\cite{kolve2022} environment. The reconstructed task-specific images under Rayleigh channel are presented for different subtasks to illustrate how the proposed framework can effectively adapt to subtask-relevant regions. Simulation results demonstrate that our proposed HiTOC framework outperforms three state-of-the-art schemes in terms of the success rate on the MAP-THOR~\cite{nayak2024mapthor} benchmark for long-horizon tasks. 
\end{itemize}
This paper is organized as follows. Section~\ref{sec:framework} presents the system model of the proposed HiTOC framework. Section~\ref{sec:cVIB} introduces the cVIB objective and derive its upper bound. Section~\ref{sec:model} presents the architecture of the main modules and the training procedure of HiTOC. The performance evaluation is presented in Section~\ref{sec:evaluation}. Section~\ref{sec:conclude} concludes the paper.


\textit{Notation:} 
The Kullback--Leibler divergence between two distributions $p(\mathbf{x})$ and $q(\mathbf{x})$ is denoted by $D_{\mathrm{KL}}(p(\mathbf{x})\,\|\,q(\mathbf{x}))$. The mutual information between two random variables $X$ and $Y$ is denoted by $I(X;Y)$, and the conditional mutual information given $Z$ is denoted by $I(X;Y |Z)$. The entropy of $X$ is denoted by $H(X)$, and the conditional entropy given $Y$ is denoted by $H(X|Y)$. The real Gaussian and complex Gaussian distributions with mean $\boldsymbol{\mu}$ and covariance $\boldsymbol{\Sigma}$ are denoted by $\mathcal{N}(\boldsymbol{\mu}, \boldsymbol{\Sigma})$ and $\mathcal{CN}(\boldsymbol{\mu}, \boldsymbol{\Sigma})$, respectively. The zero vector and identity matrix are denoted by $\mathbf{0}$ and $\mathbf{I}$, respectively.

\section{The Proposed Hierarchical Task-Oriented Communication Framework}
\label{sec:framework}
\begin{figure}[t]
    \includegraphics[width=\linewidth]{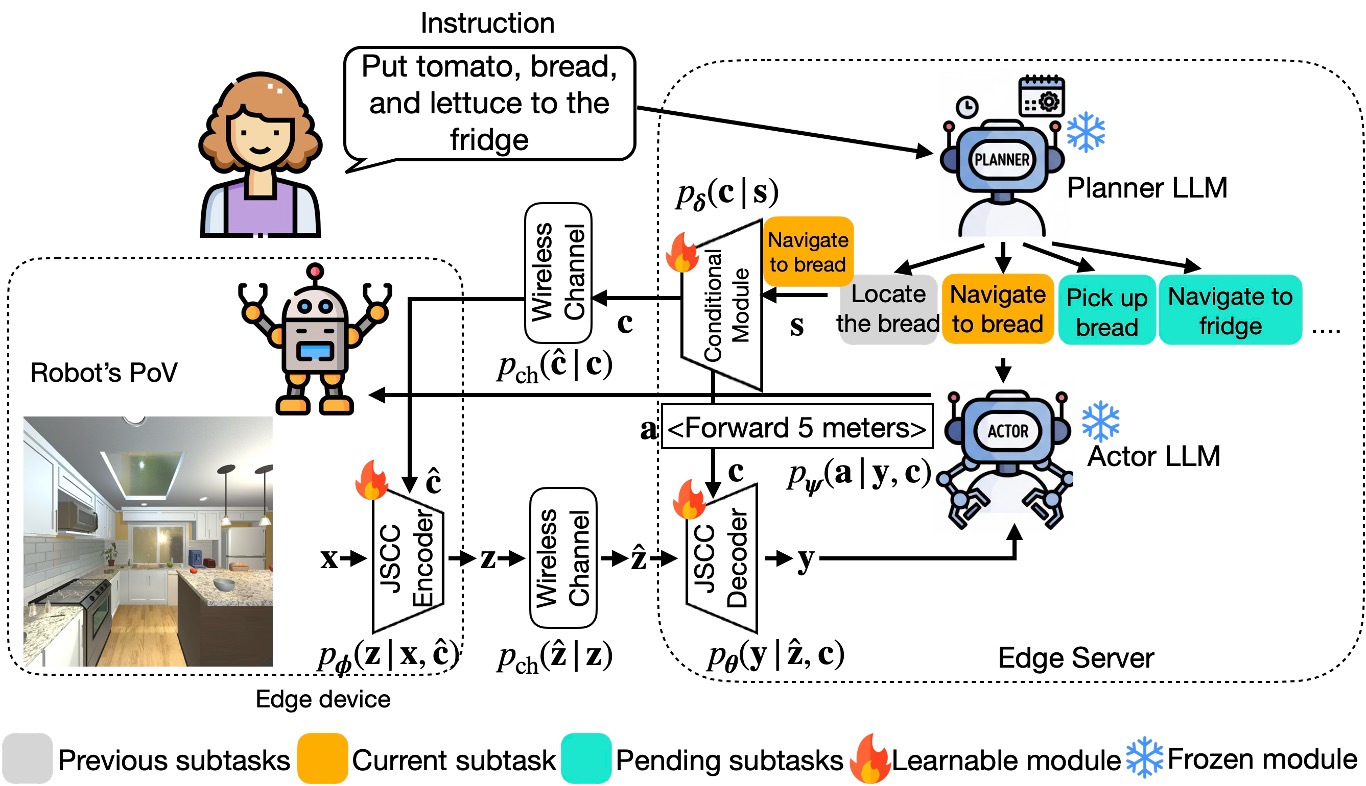}
    \caption{The system model of our proposed HiTOC framework.}
    \label{fig:system_model}
    \vspace{-20pt}
\end{figure}
The system model of our proposed HiTOC framework is shown in Fig.~\ref{fig:system_model}. It consists of an edge server and an edge device. The edge server includes a planner and an actor LLM, while the edge device corresponds to a robot controlled by the edge server. When a human user provides a task instruction to the edge server, the planner LLM module reasons about the instruction and decomposes the task into a sequence of subtasks. The subtasks are processed sequentially. Each subtask may require multiple actions to be completed. For example, the subtask “locate the bread” may require the robot to rotate 30 degrees multiple times to search for the bread. The actions for each subtask are determined by the actor LLM module. To choose an action, the actor module requires updated observations from the environment as input.

We denote each subtask generated by the planner LLM as a random variable $S$ over the subtask space $\mathcal{S}$. The current subtask $\mathbf{s}$ is sent to the conditional module at the edge server, which encodes the subtask description into a latent representation denoted by $\mathbf{c}\in \mathbb{C}^{K}$, where $K$ is the dimension of latent space. Then, the vector $\mathbf{c}$ is transmitted through the wireless channel to the edge device to provide information about the current subtask. The received subtask representation is denoted as $\hat{\mathbf{c}} \in \mathbb{C}^{K}$. For a Rayleigh fading channel, the received subtask representation $\hat{\mathbf{c}}$ can be expressed as
\begin{align}
    \hat{\mathbf{c}} = h_{c}\mathbf{c}+\mathbf{n}_c,
    \label{eq:channel_c}
\end{align}
where $h_{c} \sim \mathcal{CN}(0,1)$ is the channel fading gain and $\mathbf{n}_c \sim \mathcal{CN}(0, \sigma_{c}^{2}\mathbf{I})$ is the additive white Gaussian noise (AWGN) with $\sigma_{c}^2$ being the noise variance. To provide the updated observation from the environment, the robot captures a point-of-view (PoV) image $\mathbf{x}\in\mathbb{R}^{H\times W\times D}$, where $H$, $W$, and $D$ are the height, width, and number of channels for the image, respectively. The image $\mathbf{x}$ is encoded by the JSCC encoder to generate the latent representation $\mathbf{z}\in \mathbb{C}^{K}$. After transmission through the wireless channel, the received latent representation, denoted by $\hat{\mathbf{z}}\in \mathbb{C}^{K}$, is given by
\begin{align}
    \hat{\mathbf{z}} = h_{z}\mathbf{z}+\mathbf{n}_{z},
\end{align}
where $h_{z} \sim \mathcal{CN}(0,1)$ is the channel fading gain and $\mathbf{n}_z \sim \mathcal{CN}(0, \sigma_{z}^{2}\mathbf{I})$ is the AWGN with $\sigma_{z}^2$ being the noise variance. 
Since the received latent representation $\hat{\mathbf{z}}$ does not match the input dimensionality required by the actor LLM module, the JSCC decoder acts as an information reshaper that transforms $\hat{\mathbf{z}}$ into a task-specific image $\mathbf{y}\in\mathbb{R}^{H\times W\times D}$. Similar reshaping mechanism was also used in~\cite{Diao2025}. The actor LLM then chooses an action $\mathbf{a}$ from the action space $\mathcal{A}$ based on $\mathbf{y}$ and $\mathbf{c}$.

\begin{figure}[t]
    \centering
    \includegraphics[width=0.6\linewidth]{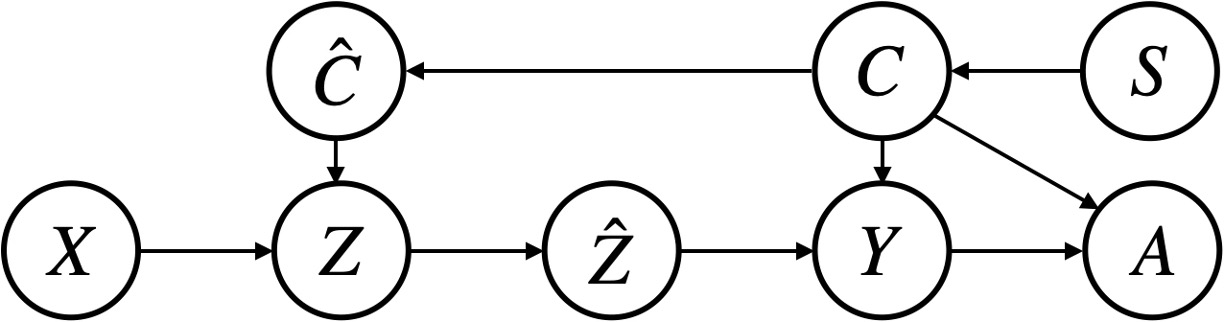}
    \caption{The probabilistic graphical model of the proposed HiTOC framework.}
    \vspace{-20pt}
    \label{fig:graph}
\end{figure}

In our framework, the subtask representation serves as side information for the JSCC encoding and decoding processes. The side information aligns the encoded representation $\mathbf{z}$ and task-specific image $\mathbf{y}$ with the subtask context. The probabilistic graphical model of our proposed framework is illustrated as Fig.~\ref{fig:graph}.
The chain $S\rightarrow C\rightarrow \hat{C}$ represents the generation and transmission of the subtask latent representation, while the chain $X\rightarrow Z\rightarrow \hat{Z}\rightarrow Y\rightarrow A$ represents the encoding and decoding processes. The encoder is conditioned on the received subtask latent representation $\hat{C}$, as it operates on the edge device, whereas the decoder at the edge server uses the original representation $C$.
Let the conditional module be parameterized by $\boldsymbol{\delta}$. The distribution of the subtask representation $\mathbf{c}$ given a subtask $\mathbf{s}$ is defined as $p_{\boldsymbol{\delta}}(\mathbf{c}\,|\, \mathbf{s})$. The JSCC encoder, parameterized by $\boldsymbol{\phi}$, defines the distribution of the latent representation $\mathbf{z}$ as $p_{\boldsymbol{\phi}}(\mathbf{z}\,|\, \mathbf{x}, \hat{\mathbf{c}})$, conditioned on the input image $\mathbf{x}$ and the received subtask representation $\hat{\mathbf{c}}$. The JSCC decoder, parameterized by $\boldsymbol{\theta}$, defines the distribution of the task-specific image as $p_{\boldsymbol{\theta}}(\mathbf{y}\,|\, \hat{\mathbf{z}}, \mathbf{c})$, conditioned on the received latent representation $\hat{\mathbf{z}}$ and the subtask variable $\mathbf{c}$. The actor LLM, parameterized by $\boldsymbol{\psi}$, defines the distribution of the actions as 
$p_{\boldsymbol{\psi}}(\mathbf{a}\,|\, \mathbf{y}, \mathbf{c})$, given the task-specific image $\mathbf{y}$ and the subtask variable $\mathbf{c}$. $p_{\text{ch}}(\hat{\mathbf{z}}\,|\, \mathbf{z})$ and $p_{\text{ch}}(\hat{\mathbf{c}}\,|\, \mathbf{c})$ define the transmission of the latent features and the subtask representation, respectively. Based on the graphical model, the joint distribution of all variables is given by
\begin{align}
&~p(\mathbf{x}, \mathbf{z}, \hat{\mathbf{z}}, \mathbf{y}, \mathbf{a}, \mathbf{c}, \hat{\mathbf{c}}, \mathbf{s}) \nonumber\\
&= p(\mathbf{s})\; p_{\boldsymbol{\delta}}(\mathbf{c}\mid \mathbf{s}) \;
   p_{\text{ch}}(\hat{\mathbf{c}}\mid \mathbf{c}) \;
   p(\mathbf{x}, \mathbf{z}, \hat{\mathbf{z}}, \mathbf{y}, \mathbf{a}\mid \hat{\mathbf{c}}, \mathbf{c}),
\end{align}
where $p(\mathbf{x}, \mathbf{z}, \hat{\mathbf{z}}, \mathbf{y}, \mathbf{a}\mid \hat{\mathbf{c}}, \mathbf{c})$ can be factorized as
\begin{align}
&~p(\mathbf{x}, \mathbf{z}, \hat{\mathbf{z}}, \mathbf{y}, \mathbf{a}\mid \hat{\mathbf{c}}, \mathbf{c}) \nonumber\\
&= p(\mathbf{x}) \;
  p_{\boldsymbol{\phi}}(\mathbf{z}\mid \mathbf{x}, \hat{\mathbf{c}}) \;
  p_{\text{ch}}(\hat{\mathbf{z}}\mid \mathbf{z}) \;
  p_{\boldsymbol{\theta}}(\mathbf{y}\mid \hat{\mathbf{z}}, \mathbf{c}) \;
  p_{\boldsymbol{\psi}}(\mathbf{a}\mid \mathbf{y}, \mathbf{c}).
  \label{eq:p(...|c^)}
\end{align}
Note that the environment observation and the channel are independent of the subtask; hence, $p(\mathbf{x}\mid\hat{\mathbf{c}},\mathbf{c})=p(\mathbf{x})$ and $p_{\text{ch}}(\hat{\mathbf{z}}\mid \mathbf{z}, \hat{\mathbf{c}},\mathbf{c})=p_{\text{ch}}(\hat{\mathbf{z}}\mid \mathbf{z})$. 

Here, direct computation of conditional distribution $p_{\boldsymbol{\phi}}(\mathbf{z}\,|\, \mathbf{x},\hat{\mathbf{c}})$ 
is generally intractable since $\mathbf{z}$ has a high-dimensional dependence on both $\mathbf{x}$ and $\hat{\mathbf{c}}$. To obtain a tractable approximation, we first model the unconditional base distribution $p_{\boldsymbol{\phi}}(\mathbf{z}\,|\, \mathbf{x})$. We then introduce a variational distribution $q(\mathbf{z}\,|\, \mathbf{x}, \hat{\mathbf{c}})$ to approximate the intractable conditional distribution $p_{\boldsymbol{\phi}}(\mathbf{z}\,|\, \mathbf{x}, \hat{\mathbf{c}})$. The variational distribution $q(\mathbf{z}\,|\, \mathbf{x}, \hat{\mathbf{c}})$ should jointly align with the subtask latent representation while remaining close to the base distribution $p_{\boldsymbol{\phi}}(\mathbf{z}\,|\, \mathbf{x})$. Specifically, we formulate the following variational problem:
\begin{align}
\min_{q} 
-\mathbb{E}_{q(\mathbf{z}|\mathbf{x},\hat{\mathbf{c}})} \big[ R(\mathbf{z},\hat{\mathbf{c}}) \big] 
+ \tfrac{1}{\gamma}\, 
D_{\mathrm{KL}}\!\big(q(\mathbf{z}\,|\, \mathbf{x},\hat{\mathbf{c}})\,\|\,p_{\boldsymbol{\phi}}(\mathbf{z}\,|\, \mathbf{x})\big), \nonumber
\end{align}
where $\gamma$ is a positive scaling parameter and $R(\mathbf{z},\hat{\mathbf{c}})$ is a nonnegative similarity function that measures the similarity between $\mathbf{z}$ and the subtask representation variable $\hat{\mathbf{c}}$.
The optimal solution of this variational problem corresponds to an exponentially tilted approximation~\cite[Th.~11.4.1]{Cover2006}, which softly tilts the base distribution toward the subtask representation $\hat{\mathbf{c}}$. Hence, we approximate $p_{\phi}(\mathbf{z}\,|\, \mathbf{x},\hat{\mathbf{c}})$ 
with an exponentially tilted distribution $\tilde{q}_{\boldsymbol{\phi}}(\mathbf{z}\,|\, \mathbf{x},\hat{\mathbf{c}})$ as follows
\begin{align}
\tilde{q}_{\boldsymbol{\phi}}(\mathbf{z}\,|\, \mathbf{x},\hat{\mathbf{c}}) 
= \frac{p_{\boldsymbol{\phi}}(\mathbf{z}\,|\, \mathbf{x})\,
\exp\!\big[\gamma R(\mathbf{z},\hat{\mathbf{c}})\big]}
{Z_\gamma(\mathbf{x},\hat{\mathbf{c}})},
\label{eq:p_soft}
\end{align}
where $Z_\gamma(\mathbf{x},\hat{\mathbf{c}})$ is the normalizing function defined as
\begin{align}
Z_\gamma(\mathbf{x},\hat{\mathbf{c}}) 
= \int p_{\boldsymbol{\phi}}(\mathbf{z}\,|\, \mathbf{x})\,
   \exp\!\big[\gamma R(\mathbf{z},\hat{\mathbf{c}})\big]\,
   d\mathbf{z}.
\end{align}
When $\gamma \rightarrow 0$, the tilted distribution $\tilde{q}_{\boldsymbol{\phi}}(\mathbf{z}\,|\, \mathbf{x},\hat{\mathbf{c}})$ converges to the base encoder distribution $p_{\boldsymbol{\phi}}(\mathbf{z}\,|\,\mathbf{x})$, meaning that the encoding no longer depends on the subtask. Similarly, we apply exponential tilting to approximate $p_{\boldsymbol{\theta}}(\mathbf{y}\,|\, \hat{\mathbf{z}},\mathbf{c})$, which is denoted as $\tilde{q}_{\boldsymbol{\theta}}(\mathbf{y}\,|\, \hat{\mathbf{z}},\mathbf{c})$. With the exponentially tilted approximation, $p(\mathbf{x}, \mathbf{z}, \hat{\mathbf{z}}, \mathbf{y}, \mathbf{a}\,|\,\hat{\mathbf{c}}, \mathbf{c})$ in~\eqref{eq:p(...|c^)} can be approximated as
\begin{align}
&~p(\mathbf{x}, \mathbf{z}, \hat{\mathbf{z}}, \mathbf{y}, \mathbf{a}\mid \hat{\mathbf{c}}, \mathbf{c}) \nonumber\\
&\approx p(\mathbf{x}) \;
  \tilde{q}_{\boldsymbol{\phi}}(\mathbf{z}\mid \mathbf{x}, \hat{\mathbf{c}}) \;
  p_{\text{ch}}(\hat{\mathbf{z}}\mid \mathbf{z}) \;
  \tilde{q}_{\boldsymbol{\theta}}(\mathbf{y}\mid \hat{\mathbf{z}}, \mathbf{c}) \;
  p_{\boldsymbol{\psi}}(\mathbf{a}\mid \mathbf{y}, \mathbf{c}).
   \label{eq:tildep(...|c^)no_c}
\end{align}

\section{The conditional variational information bottleneck approach }
\label{sec:cVIB}
In the proposed HiTOC framework, we consider the following three objectives: (i) train a conditional module that encodes the subtask representation $\mathbf{c}$ to preserve the action-relevant information from the subtask $\mathbf{s}$; (ii) train the JSCC encoder and decoder such that the encoder produces $\mathbf{z}$ given $\hat{\mathbf{c}}$ and the decoder reconstructs $\mathbf{y}$ given $\mathbf{c}$, while only keeping information related to the action; and (iii) ensure that the task-specific image $\mathbf{y}$ retains the action-related information contained in the received latent representation $\hat{\mathbf{z}}$. To jointly achieve these objectives, we define the following loss function:
\begin{align}
\mathcal{L}_{\text{cIB}}(\mathbf{a},\mathbf{x},\mathbf{s};\boldsymbol{\phi}, \boldsymbol{\theta},& \boldsymbol{\delta}) 
= \Big[ \underbrace{- I(A;Y \mid C)}_{\text{Task data distortion}} 
+ \beta_1 \underbrace{I(X;\hat{Z} \mid \hat{C})}_{\text{Task data rate}} \Big] \nonumber \\
&~ + \beta_2 \Big[ \underbrace{I(A;Y \mid C) - I(A;\hat{Z} \mid C)}_{\text{Alignment}} \Big] \nonumber \\
+  &~\Big[ \beta_3 \underbrace{I(S;C)}_{\text{Subtask rate}}- \beta_4\underbrace{I(A;C\mid Y)}_{\text{Subtask data distortion}}  \Big],
\label{eq:obj}
\end{align}
where $\beta_1, \beta_3$ and $\beta_4$ are positive regularization weights for minimizing the rate of task data $I(X;\hat{Z}\mid \hat{C})$, subtask data rate $I(S;C)$ and subtask data distortion $I(A;C\mid Y)$ , respectively. We restrict $\beta_2$ to be $0<\beta_2 <1$ so that the coefficient of $I(A;Y \mid \hat{C})$ remains negative to maximize the information transfer from $Y$ to action $A$. The term in the first bracket, $- I(A;Y \mid C)+\beta_1I(X;\hat{Z} \mid \hat{C})$, forms the information bottleneck on task data~\cite{Shao2022} conditioned on $C$ and $\hat{C}$, which addresses objective (ii). The alignment term, $I(A;Y \mid C) - I(A;\hat{Z} \mid C)$, ensures $\hat{Z}$ and $Y$ retain the same information about the action $A$~\cite{Diao2025}, which addresses objective (iii). The term in the last bracket, $\beta_3I(S;C)-\beta_4I(A;C\mid Y)$, is the information bottleneck for subtask representation, which addresses objective (i). We rescale~\eqref{eq:obj} by $(1-\beta_2)$:
\begin{align}
\mathcal{L}_{\text{cIB}}(\mathbf{a},\mathbf{x},\mathbf{s};\boldsymbol{\phi}, \boldsymbol{\theta}, \boldsymbol{\delta}) 
&=- I(A;Y \mid C) + \hat{\beta_1} I(X;\hat{Z} \mid \hat{C}) \nonumber \\
- \hat{\beta_2}I(A;\hat{Z} \mid C)& + \hat{\beta_3} I(S;C)- \hat{\beta_4} I(A;C\mid Y), \label{eq:lag}
\end{align}
where $\hat{\beta}_i = \beta_i/(1 - \beta_2)$ for $i\in \{1, 2, 3, 4\}$.

We next derive the upper bounds for each term in~\eqref{eq:lag}, starting with the term $I(X;\hat{Z}\mid \hat{C})$. Specifically,  
\begin{align}
\label{eq:I(X;Zh|Ch)}
I(X;\hat{Z}\mid \hat{C}) 
= \mathbb{E}_{\mathbf{\hat{c}}} \; \mathbb{E}_{\mathbf{x}} 
\Big[ D_{\mathrm{KL}}\!\big(p_{\boldsymbol{\phi}}(\hat{\mathbf{z}}\mid \mathbf{x},\mathbf{\hat{c}})\,\|\,p(\hat{\mathbf{z}}\mid \mathbf{\hat{c}})\big) \Big].
\end{align}
\noindent Here, $p_{\boldsymbol{\phi}}(\hat{\mathbf{z}}\,|\, \mathbf{x},\hat{\mathbf{c}}) = p_{\text{ch}}(\hat{\mathbf{z}}\,|\,\mathbf{z})\tilde{q}_{\boldsymbol{\phi}}(\mathbf{z}\,|\, \mathbf{x},\hat{\mathbf{c}})$ denotes the conditional distribution produced by the encoder and channel, where $\tilde{q}_{\boldsymbol{\phi}}(\mathbf{z}\,|\,\mathbf{x},\hat{\mathbf{c}})$ is the exponential tilt approximation of the conditional encoder distribution given in~\eqref{eq:p_soft}. The marginal distribution of  $p(\hat{\mathbf{z}}\,|\, \hat{\mathbf{c}})$ is given by
\begin{align}
p(\hat{\mathbf{z}}\mid \hat{\mathbf{c}}) = \int p(\mathbf{x})\, \tilde{q}_{\boldsymbol{\phi}}(\mathbf{z}\mid \mathbf{x},\hat{\mathbf{c}})\, p_{\text{ch}}(\hat{\mathbf{z}}\mid \mathbf{z})\,d\mathbf{z}d\mathbf{x},
\end{align}
which is intractable.
To address this, 
we approximate $p(\hat{\mathbf{z}}\,|\, \hat{\mathbf{c}})$ with a tractable prior $q(\hat{\mathbf{z}}\,|\, \hat{\mathbf{c}})$. Since $\mathbb{E}_{\hat{\mathbf{c}}}\Big[D_{\mathrm{KL}}\!\big(p(\hat{\mathbf{z}}\,|\, \hat{\mathbf{c}})\,\|\,q(\hat{\mathbf{z}}\,|\, \hat{\mathbf{c}})\big)\Big]\geq0
$, we have
\begin{align}
I(X;\hat{Z}\mid \hat{C}) 
\leq \mathbb{E}_{\hat{\mathbf{c}}}\; \mathbb{E}_{\mathbf{x}} 
\Big[ D_{\mathrm{KL}}\!\big(p_{\boldsymbol{\phi}}(\hat{\mathbf{z}}\mid \mathbf{x},\hat{\mathbf{c}})\,\|\,q(\hat{\mathbf{z}}\mid \hat{\mathbf{c}})\big) \Big].
\label{eq:I(X;Z^|C)-ineq}
\end{align}
Next, the term $-I(A;Y\mid C)$ can be expressed as
\begin{align}
&~- I(A;Y\mid C) \nonumber \\
&= - \mathbb{E}_{\mathbf{c}} \Big[
   \int p(\mathbf{a},\mathbf{y}\mid \mathbf{c})
   \log \frac{p(\mathbf{a}\mid \mathbf{y},\mathbf{c})}{p(\mathbf{a}\mid \mathbf{c})}\,
   d\mathbf{a}\,d\mathbf{y}  \Big]\nonumber\\
&= -  \mathbb{E}_{\mathbf{c}} \Big[
   \int p(\mathbf{a},\mathbf{y}\mid \mathbf{c})
   \log p(\mathbf{a}\mid \mathbf{y},\mathbf{c})\,
   d\mathbf{a}\,d\mathbf{y}\Big] 
   - H(A\mid C),\nonumber\\
&\leq -  \mathbb{E}_{\mathbf{c}} \Big[
   \int p(\mathbf{a},\mathbf{y}\mid \mathbf{c})
   \log p(\mathbf{a}\mid \mathbf{y},\mathbf{c})\,
   d\mathbf{a}\,d\mathbf{y}\Big] \label{eq:I(A;Y|C)-3},
\end{align}
where $p(\mathbf{a}\mid \mathbf{y},\mathbf{c})$ denotes the posterior distribution of $A$ given $Y$ and $C$. Since $H(A\mid C)\geq 0$, removing this term yields the upper bound in~\eqref{eq:I(A;Y|C)-3}. The posterior $p(\mathbf{a}\mid \mathbf{y},\mathbf{c})$ can be derived from the joint distribution in~\eqref{eq:tildep(...|c^)no_c} as
\begin{align}
&~ p(\mathbf{a}\mid \mathbf{y}, \mathbf{c})=  \nonumber \\
&\!\!\int\!\frac{p(\mathbf{x})
   \tilde{q}_{\boldsymbol{\phi}}(\mathbf{z}| \mathbf{x},\!\mathbf{c})
   p_{\text{ch}}(\hat{\mathbf{z}}|\mathbf{z})
   \tilde{q}_{\boldsymbol{\theta}}(\mathbf{y}\,| \hat{\mathbf{z}},\!\mathbf{c})
   p_{\boldsymbol{\psi}}(\mathbf{a}\,|\mathbf{y},\!\mathbf{c})}
   {p(\mathbf{y}\,|\,\mathbf{c})}d\mathbf{x}d\mathbf{z}d\hat{\mathbf{z}},
\label{eq:p(a|y,c)}
\end{align}
where $\tilde{q}_{\boldsymbol{\phi}}(\mathbf{z}| \mathbf{x},\mathbf{c}) \triangleq \int p_{\text{ch}}(\hat{\mathbf{c}}|\mathbf{c})\tilde{q}_{\boldsymbol{\phi}}(\mathbf{z}|\mathbf{x}, \hat{\mathbf{c}})d\hat{\mathbf{c}}$ denotes the tilted encoder distribution marginalized over the channel. 
Since~\eqref{eq:p(a|y,c)} is intractable,  
we introduce a variational posterior $q_{\boldsymbol{\psi}}(\mathbf{a}\mid \mathbf{y},\mathbf{c})$ to approximate $p(\mathbf{a}| \mathbf{y},\mathbf{c})$. Based on the fact that
$D_{\mathrm{KL}}\!\big(p(\mathbf{a}\mid \mathbf{y},\mathbf{c}) \,\|\, q_{\boldsymbol{\psi}}(\mathbf{a}\mid \mathbf{y},\mathbf{c})\big)\geq0$, we obtain
\begin{align}
-\mathbb{E}_{\mathbf{c}} \Big[\int p(\mathbf{a}, \mathbf{y} \mid \mathbf{c}) 
    \log p(\mathbf{a} \mid \mathbf{y}, \mathbf{c}) \, d\mathbf{a}\, d\mathbf{y}\Big]
\;\; \nonumber\\
\leq-\mathbb{E}_{\mathbf{c}} \Big[\int p(\mathbf{a}, \mathbf{y} \mid \mathbf{c}) 
    \log q_{\boldsymbol{\psi}}(\mathbf{a} \mid \mathbf{y}, \mathbf{c}) \, d\mathbf{a}\, d\mathbf{y}\Big].
\label{eq:int-ineq}
\end{align}
Hence,
\begin{align}
- I(A;Y \mid C)
\leq
\mathbb{E}_{\mathbf{c}}
  \mathbb{E}_{\hat{\mathbf{c}} \mid \mathbf{c}}
    \mathbb{E}_{\mathbf{a},\mathbf{x}}    \mathbb{E}_{\mathbf{y}|\mathbf{x},\mathbf{c},\hat{\mathbf{c}};\boldsymbol{\phi},\boldsymbol{\theta}}
      \big[
        - \log q_{\boldsymbol{\psi}}(\mathbf{a} \mid \mathbf{y}, \mathbf{c})
      \big].
\label{eq:I(A;Y|C)-up}
\end{align}
Similar as~\eqref{eq:I(A;Y|C)-3}, the term $I(A;\hat{Z}\mid C)$ can be written as
\begin{align}
&~I(A;\hat{Z}\mid C) \nonumber\\
&= \mathbb{E}_{\mathbf{c}} \Big[\int p(\mathbf{a},\hat{\mathbf{z}}\mid \mathbf{c}) 
   \log p(\mathbf{a}\mid\hat{\mathbf{z}}, \mathbf{c})\,
   d\mathbf{a}\,d\hat{\mathbf{z}}\Big] - H(A\mid C) \nonumber\\
&\leq \mathbb{E}_{\mathbf{c}} \Big[\int p(\mathbf{a},\hat{\mathbf{z}}\mid \mathbf{c}) 
   \log p(\mathbf{a}\mid\hat{\mathbf{z}}, \mathbf{c})\,
   d\mathbf{a}\,d\hat{\mathbf{z}}\Big]  \label{eq:I(A;Z^|C)-3}
\end{align}
where $p(\mathbf{a}\mid \hat{\mathbf{z}},\mathbf{c})$ denotes the posterior of $A$ given $\hat{Z}$ and $C$. We again use the fact $H(A\mid C)\geq0$ to obtain the upper bound in~\eqref{eq:I(A;Z^|C)-3}. Following the conditional joint distribution~\eqref{eq:tildep(...|c^)no_c}, the posterior $p(\mathbf{a}\mid \hat{\mathbf{z}},\mathbf{c})$ can be expressed as
\begin{align}
&~p(\mathbf{a}\mid \hat{\mathbf{z}}, \mathbf{c}) = \nonumber\\
&\!\!\int\! \frac{p(\mathbf{x})
   \tilde{q}_{\boldsymbol{\phi}}(\mathbf{z}| \mathbf{x}, \mathbf{c})
   p_{\text{ch}}(\hat{\mathbf{z}}| \mathbf{z})
   \tilde{q}_{\boldsymbol{\theta}}(\mathbf{y}| \hat{\mathbf{z}}, \mathbf{c})
   p_{\boldsymbol{\psi}}(\mathbf{a}| \mathbf{y}, \mathbf{c})}
     {p(\hat{\mathbf{z}}\,|\, \mathbf{c})}d\mathbf{x}d\mathbf{z}d\mathbf{y},
\end{align}
where $\tilde{q}_{\boldsymbol{\phi}}(\mathbf{z}\,|\,\mathbf{x},\mathbf{c})$ is the tilted encoder distribution marginalized over the channel, as used in~\eqref{eq:p(a|y,c)}.
Due to the intractability of $p(\mathbf{a}\,|\, \hat{\mathbf{z}}, \mathbf{c})$, we introduce a variational distribution $q_{\boldsymbol{\psi}}(\mathbf{a}\,|\,\hat{\mathbf{z}},\mathbf{c})$ to approximate $p(\mathbf{a}\,|\, \hat{\mathbf{z}},\mathbf{c})$.  
By definition, we have $D_{\mathrm{KL}}\!\big(p(\mathbf{a}\,|\,\hat{\mathbf{z}},\mathbf{c}) 
 \,\|\, q_{\boldsymbol{\psi}}(\mathbf{a}\,|\, \hat{\mathbf{z}},\mathbf{c})\big) \geq 0.$
Similar as~\eqref{eq:int-ineq}, we obtain
\begin{align}
- I(A;\hat{Z} | C)
\leq
\mathbb{E}_{\mathbf{c}}
  \mathbb{E}_{\hat{\mathbf{c}} | \mathbf{c}}
    \mathbb{E}_{\mathbf{a},\mathbf{x}}
      \mathbb{E}_{\hat{\mathbf{z}}|\mathbf{x}, \hat{\mathbf{c}};\boldsymbol{\phi}}
      \big[
        - \log q_{\boldsymbol{\psi}}(\mathbf{a} | \hat{\mathbf{z}}, \mathbf{c})
      \big].
\label{eq:I(A;Z^|C)-up}
\end{align}
Next, the term $I(S;C)$ can be expressed as
\begin{align}
I(S;C)
= \mathbb{E}_{\mathbf{s}}
\Big[ D_{\mathrm{KL}}\!\big(p_{\boldsymbol{\delta}}(\mathbf{c}\mid \mathbf{s})\,\|\,p(\mathbf{c})\big) \Big].
\label{eq:I(S;C)-exact}
\end{align}
\noindent
The true marginal $p(\mathbf{c})=\int p_{\boldsymbol{\delta}}(\mathbf{c}\mid \mathbf{s})\,p(\mathbf{s})\,d\mathbf{s}$ is generally intractable. We also
introduce $q(\mathbf{c})\sim\mathcal{N}(\mathbf{0},\mathbf{I})$ as a tractable prior for $p(\mathbf{c})$. With $D_{\mathrm{KL}}\!\big(p(\mathbf{c})\,\|\,q(\mathbf{c})\big)\!\ge\!0$, we obtain
\begin{align}
I(S;C)\le \mathbb{E}_{\mathbf{s}} \Big[ D_{\mathrm{KL}}\!\big(p_{\boldsymbol{\delta}}(\mathbf{c}\mid \mathbf{s})\,\|\,q(\mathbf{c})\big) \Big].
\label{eq:I(S;C)-upper}
\end{align}
Similar as~\eqref{eq:I(A;Y|C)-3}, the upper bound of $-I(A;C\mid Y)$ is given by
\begin{align}
-I(A;C\mid Y)
\leq -\mathbb{E}_{\mathbf{y}}\!\Big[ \int p(\mathbf{a}, \mathbf{c}\mid\mathbf{y})
\log p(\mathbf{a}\mid \mathbf{c},\mathbf{y})\, d\mathbf{a}d\mathbf{c} \Big]. \label{eq:I(A;C|Y)-3}
\end{align}
With the variational posterior $q_{\boldsymbol\psi}(\mathbf{a}\mid \mathbf{c},\mathbf{y})$, we have the upper bound same as~\eqref{eq:I(A;Y|C)-up} by applying law of total expectation:
\begin{align}
-\,I(A&;C\mid Y)
\leq\,\mathbb{E}_{\mathbf{c}}\mathbb{E}_{\hat{\mathbf{c}}|\mathbf{c}}\mathbb{E}_{\mathbf{a},\mathbf{x}}\mathbb{E}_{\mathbf{y}\mid \mathbf{x},\mathbf{c},\hat{\mathbf{c}};\,\boldsymbol{\phi},\boldsymbol{\theta}} 
   \big[-\log q_{\boldsymbol\psi}(\mathbf{a}\mid \mathbf{c},\mathbf{y})\big].
   \label{eq:I(A;C|Y)-up}
\end{align}
Based on the upper bound we obtained in~\eqref{eq:I(X;Z^|C)-ineq},~\eqref{eq:I(A;Y|C)-up},~\eqref{eq:I(A;Z^|C)-up},~\eqref{eq:I(S;C)-upper} and~\eqref{eq:I(A;C|Y)-up}, we derive the upper bound of the loss function in~\eqref{eq:obj} as
\begin{align}
&\mathcal{L}_{\text{cVIB}}(\mathbf{a},\mathbf{x},\mathbf{s}; \boldsymbol{\phi}, \boldsymbol{\theta}, \boldsymbol{\delta}) \nonumber \\
&\leq \,\mathbb{E}_{\hat{\mathbf{c}}} \bigg[\hat{\beta}_1 \,\mathbb{E}_{\mathbf{x}} 
 \Big[ D_{\mathrm{KL}}\!\big(p_{\boldsymbol{\phi}}(\hat{\mathbf{z}}\mid \mathbf{x},\hat{\mathbf{c}}) \,\|\, q(\hat{\mathbf{z}}\mid \hat{\mathbf{c}})\big) \Big] \bigg] \nonumber\\
&\quad + \mathbb{E}_{\mathbf{c}} \mathbb{E}_{\hat{\mathbf{c}}|\mathbf{c}}\mathbb{E}_{\mathbf{a},\mathbf{x}}\bigg[ (1+\hat{\beta}_4)\mathbb{E}_{\mathbf{y}\mid \mathbf{x},\mathbf{c},\hat{\mathbf{c}};\,\boldsymbol{\phi},\boldsymbol{\theta}} 
   \big[ - \log q_{\boldsymbol{\psi}}(\mathbf{a} \mid \mathbf{y}, \mathbf{c}) \big] \nonumber\\
&\quad + \hat{\beta}_2 \,\mathbb{E}_{\hat{\mathbf{z}}\mid \mathbf{x}, \hat{\mathbf{c}};\,\boldsymbol{\phi}} 
   \big[ - \log q_{\boldsymbol{\psi}}(\mathbf{a} \mid \hat{\mathbf{z}}, \mathbf{c}) \big] \bigg]\nonumber\\
&\quad + \hat{\beta}_3 \,\mathbb{E}_{\mathbf{s}} \Big[ D_{\mathrm{KL}}\!\big(p_{\boldsymbol{\delta}}(\mathbf{c}\mid \mathbf{s})\,\|\,q(\mathbf{c})\big) \Big].
\label{eq:upper-bound}
\end{align}


\section{Model Architecture and Training Procedure}
\label{sec:model}
\begin{figure*}[t]
\vspace{-0.3em}
\begin{align}
\mathcal{L}_{\text{cVIB}}^{'}(\mathbf{a},\mathbf{x},\mathbf{s}; \boldsymbol{\phi}, \boldsymbol{\theta}, \boldsymbol{\delta})
&= \frac{1}{B}\sum_{i\in\mathcal{B}}\Bigg[
\frac{\hat{\beta}_1}{J}\sum_{j=1}^{J}
 D_{\mathrm{KL}}\!\big(p_{\boldsymbol{\phi}}(\hat{\mathbf{z}}\mid \mathbf{x}_i,\hat{\mathbf{c}}_i^{(j)})
   \,\|\, q(\hat{\mathbf{z}}\mid \hat{\mathbf{c}}_i^{(j)})\big) 
+ \frac{(1+\hat{\beta}_4)}{J}\sum_{j=1}^{J}
   \ell_{\text{sem}}(\mathbf{x}_i, \mathbf{y}_i^{(j)}, \mathbf{c}_i^{(j)}) \nonumber\\
&+ \frac{\hat{\beta}_2}{J}\sum_{j=1}^{J}
   \big[-\log q_{\boldsymbol{\psi}}(\mathbf{a}_i\mid 
   \hat{\mathbf{z}}_{i}^{(j)},\mathbf{c}_{i}^{(j)})\big] 
+ \hat{\beta}_3\,D_{\mathrm{KL}}\!\big(p_{\boldsymbol{\delta}}(\mathbf{c}\mid \mathbf{s}_i)
   \,\|\, q(\mathbf{c})\big)
\Bigg].
\label{eq:LcVIB'}
\end{align}
\vspace{-0.4em}
\hrule
\vspace{-1.5em}
\end{figure*}
\begin{figure}
    \centering
    \includegraphics[width=\linewidth]{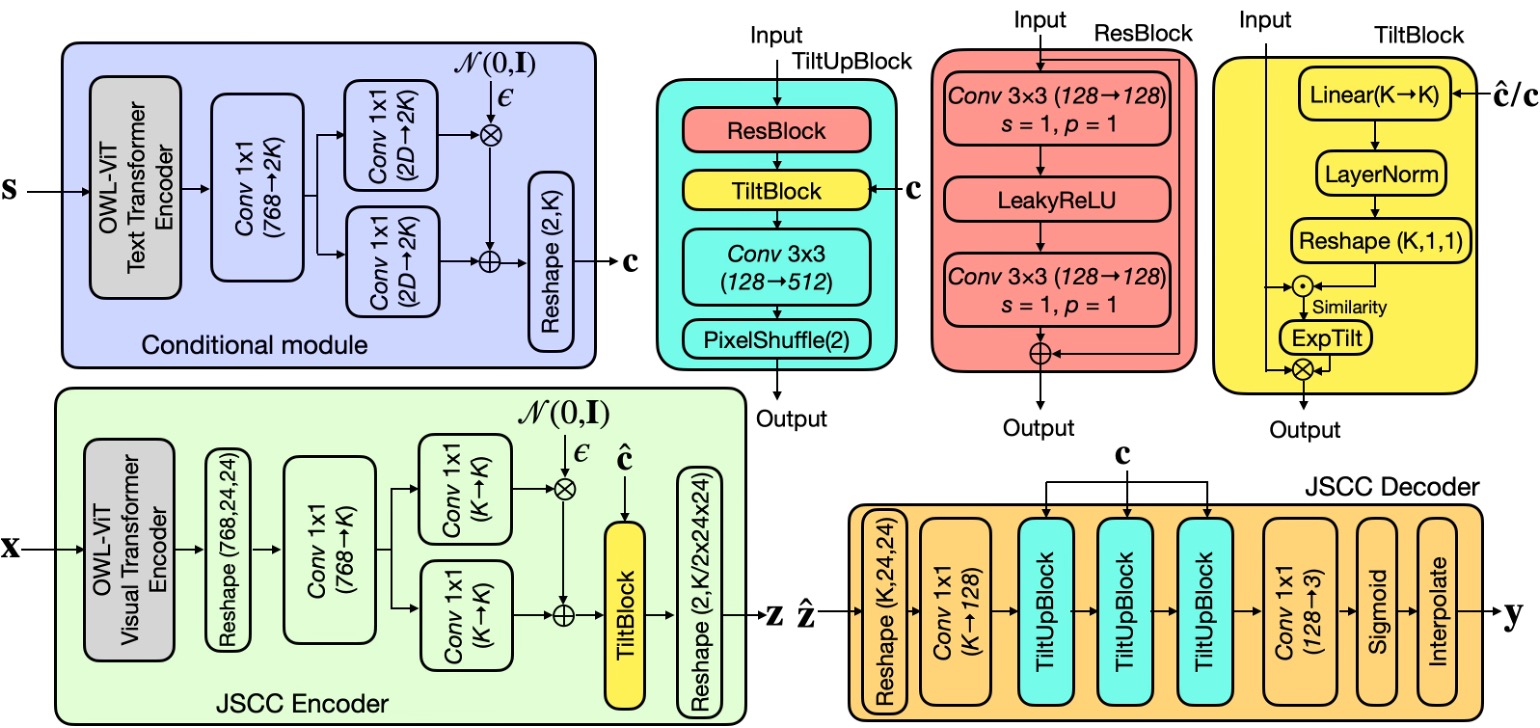}
    \caption{The architecture of the conditional module, JSCC encoder, and JSCC decoder.}
    \label{fig:model}
    \vspace{-10pt}
\end{figure}

The proposed HiTOC framework comprises three main components: the conditional module, the JSCC encoder, and the JSCC decoder. The architecture of these modules is illustrated in Fig.~\ref{fig:model}. In the figure, \textit{Conv} $1{\times}1(C_1{\rightarrow}C_2)$, $s{=}1$, $p{=}1$ denotes a $1{\times}1$ convolutional layer with stride 1 and padding 1 on both sides that maps $C_1$ to $C_2$ channels. \textit{Linear} $(C_1{\rightarrow}C_2)$ represents a fully connected layer projecting from dimension $C_1$ to $C_2$. \textit{ExpTilt}$(x)=e^{\gamma x}$ is the exponential tilt function. \textit{Reshape}$(c,h,w)$ and \textit{Reshape}$(h,w)$ denote tensor reshaping to size $(c,h,w)$ or matrix reshaping to $(h,w)$, respectively. \textit{PixelShuffle(2)} upsamples the feature map by a factor of two in height and width while reducing the channel dimension by four. The similarity between the feature and the subtask representation is measured using their inner product. One TiltBlock is used at the encoder and three in the decoder to tilt the distribution toward the subtask representation.

To train the full model, we first obtain a dataset from the simulation environment. Specifically, we use GPT-5 as both planner LLM and actor LLM, and record the image observation $\mathbf{x}_{i}$, the corresponding action $\mathbf{a}_i$, and the subtask description $\mathbf{s}_i$ for the $i$th transmission. The resulting training dataset is denoted as $\Omega=\{(\mathbf{x}_{i}, \mathbf{a}_i, \mathbf{s}_i)\}_{i=1}^{N}$, where $N$ is the total number of samples. After constructing the dataset, we employ a frozen LLaVA model as the actor LLM during training since GPT-5 is not a open source model. For each task-specific image $\mathbf{y}_i$, we concatenate the tokenized system prompt, denoted as $\text{tokenized}(P)$, with the subtask latent representation $\mathbf{c}_i$ and feed them into the LLaVA model to obtain the semantic representation $\mathbf{h}_i = \text{LLaVA}(\mathbf{y}_i, [\text{tokenized}(P),\mathbf{c}_i])$. Similarly, the reference embedding for the source image $\mathbf{x}_i$ is computed as $\tilde{\mathbf{h}}_i = \text{LLaVA}(\mathbf{x}_i, [\text{tokenized}(P),\mathbf{c}_i])$. The semantic alignment between the task-specific and source images is measured using the mean-squared error loss $\ell_{\text{sem}}(\mathbf{x}_i, \mathbf{y}_i,\mathbf{c}_i)=\|\mathbf{h}_i-\hat{\mathbf{h}}_{i}\|_2^2$. This loss serves as a differentiable surrogate for the negative log-likelihood term $-\log q_{\boldsymbol{\psi}}(\mathbf{a}_i\mid \mathbf{y}_i,\mathbf{c}_i)$ in~\eqref{eq:upper-bound}~\cite{Diao2025}.

For each mini-batch $\mathcal{B}$ of size $B$, we use Monte Carlo sampling to approximate the expectations in the cVIB objective. For each sample $j = 1, \ldots, J$, we sequentially generate $\mathbf{c}_i^{(j)}\sim p_{\boldsymbol{\delta}}(\mathbf{c}\mid \mathbf{s}_i)$ to obtain the subtask representation, transmit it through the channel to obtain $\hat{\mathbf{c}}_i^{(j)}$, then sample $\mathbf{z}_i^{(j)}\sim p_{\boldsymbol{\phi}}(\mathbf{z}\mid \mathbf{x}_i, \hat{\mathbf{c}}_i^{(j)})$ to obtain the latent features, transmit it through the channel to obtain $\hat{\mathbf{z}}_i^{(j)}$, and finally generate $\mathbf{y}_i^{(j)}\sim p_{\boldsymbol{\theta}}(\mathbf{y}_i^{(j)}\mid \hat{\mathbf{z}}_i^{(j)}, \mathbf{c}_i^{(j)})$ to obtain the task-specific image. We jointly train the JSCC encoder, decoder, and the conditional module with stochastic gradient descent while keeping LLaVA frozen. During inference, we replace LLaVA with GPT-5 for higher-level reasoning. The overall training algorithm is presented in Algorithm~\ref{alg:stage1}.
\begin{algorithm}
\small
\caption{\small Training Algorithm for the HiTOC Framework}\label{alg:stage1}
\begin{algorithmic}[1]
\State \textbf{Input:} Training dataset $\Omega:=\{(\mathbf{x}_{i}, \mathbf{a}_i, \mathbf{s}_i)\}_{i=1}^{N}$; number of epochs $E$; learning rate $\eta$; regularization weights $\hat{\beta}_{1}$, $\hat{\beta}_{2}$, $\hat{\beta}_{3}$, $\hat{\beta}_{4}$; number of samples $J$; scaling parameter $\gamma$.
\For{$e \gets 1$ \textbf{to} $E$} 
    \State Load a mini-batch $\mathcal{B} = \{(\mathbf{x}_i, \mathbf{a}_i, \mathbf{s}_i)\}_{i=1}^{B}$ from $\Omega$.
    \State Sample the channel coefficients $h_{c}$, $h_{z}$ and noise $\mathbf{n}_{c}$, $\mathbf{n}_{z}$.
    \State Perform forward propagation.
    \State Compute the loss $\mathcal{L}_{\text{cVIB}}^{'}$ based on~\eqref{eq:LcVIB'}.
    \State Update the parameters $\{\boldsymbol{\phi}, \boldsymbol{\theta}, \boldsymbol{\delta}\}$ using stochastic gradient descent with respect to the loss $\mathcal{L}_{\text{cVIB}}^{'}$.
\EndFor
\State \textbf{Output}: Optimized parameters $\{\boldsymbol{\phi}^{*}, \boldsymbol{\theta}^{*}, \boldsymbol{\delta}^{*}\}$.
\end{algorithmic}
\end{algorithm}

\section{Performance Evaluation}
\label{sec:evaluation}
We evaluate the performance of the proposed HiTOC framework and compare it with three baseline models. The conditional module and JSCC encoder are initialized using the text and visual encoders of the pretrained Owl-ViT model respectively, which has a patch size of 32 and a latent dimension of 768\footnote{Model available: \url{https://huggingface.co/google/owlvit-base-patch32}}. We project the extracted features into a latent space of dimension $K=32$. We adopt the standard normal distribution $\mathcal{N}(\mathbf{0}, \mathbf{I})$ as the tractable prior for $q(\hat{\mathbf{z}}\,|\, \hat{\mathbf{c}})$ and $q(\mathbf{c})$~\cite{Alemi2017}. We set the scaling factor $\gamma=1.1$ and the weight for each term $\hat{\beta}_1=1.0$, $\hat{\beta}_2=0.2$, $\hat{\beta}_3=1.0$, $\hat{\beta}_4=0.2$. During training, a frozen LLaVA model\footnote{Model available: \url{https://huggingface.co/liuhaotian/llava-v1.5-7b}} is used to provide differentiable semantic supervision. During inference, we employ GPT-5\footnote{Accessed via the OpenAI API: \url{https://platform.openai.com/}} to perform reasoning on the images. Experiments are conducted in the AI2-THOR~\cite{kolve2022} environment using the MAP-THOR~\cite{nayak2024mapthor} benchmark, which includes 45 household tasks across five different floor plans for each task. We compare our HiTOC framework with three baseline schemes. They are ATROC~\cite{Diao2025}, VAE~\cite{Saidu2021} and JPEG2000 with quality factor (QF) = 95~\cite{JPEG2000}. The system where the edge server receives full raw images without errors serves as the performance upper bound.

\begin{figure*}[t]
    \centering
    \begin{subfigure}{\linewidth}
        \centering
        \includegraphics[width=0.9\linewidth]{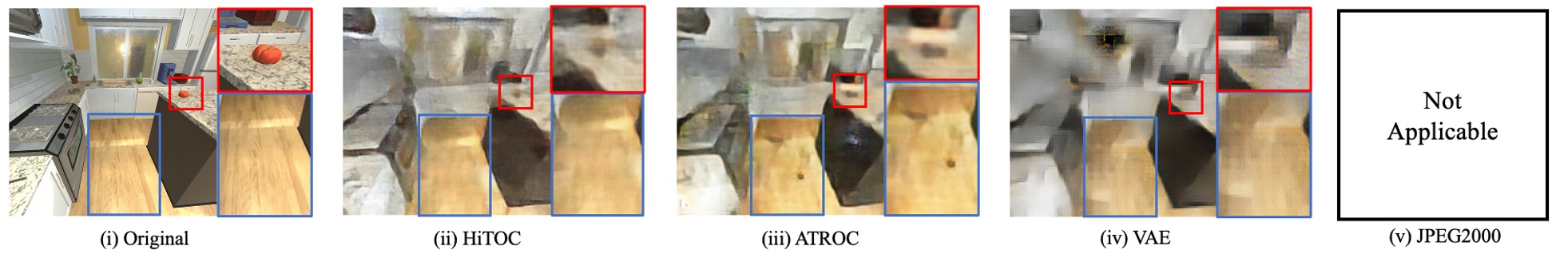}
        \vspace{-5pt}
        \caption{Task-specific image for the subtask ``Navigate to the tomato.''}
    \end{subfigure}
    \begin{subfigure}{\linewidth}
        \centering
        \includegraphics[width=0.9\linewidth]{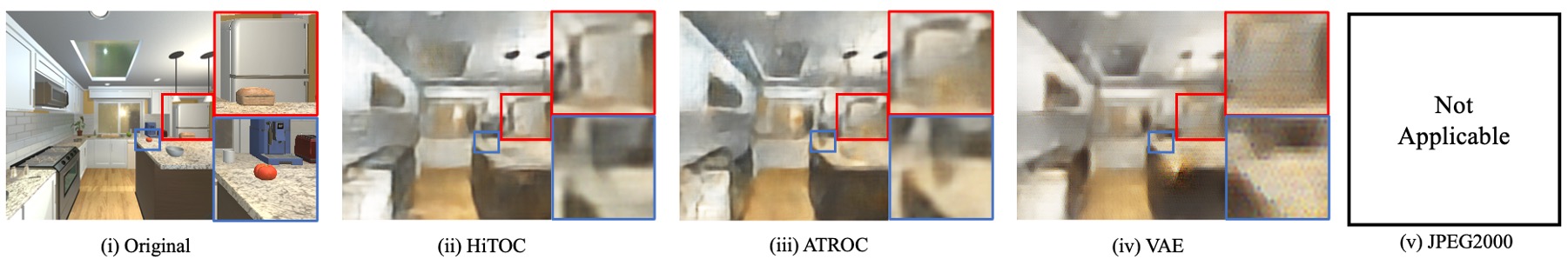}
        \vspace{-5pt}
        \caption{Task-specific image for the subtask ``Locate the fridge.''}
        \vspace{-5pt}
    \end{subfigure}
    \vspace{-10pt}
    \caption{Comparison of task-specific images from different baselines under Rayleigh fading channel with SNR = 0 dB for the subtasks of (a) ``Navigate to the tomato'' and (b) ``Locate the fridge.'' In (v), JPEG2000 is marked as \textit{Not Applicable} as it is unable to recover the image.}
    \label{fig:visualization}
    \vspace{-15pt}
\end{figure*}
\begin{figure}[t]
    \centering
    
    \begin{subfigure}[b]{0.24\textwidth}
        \centering
        \includegraphics[width=\linewidth]{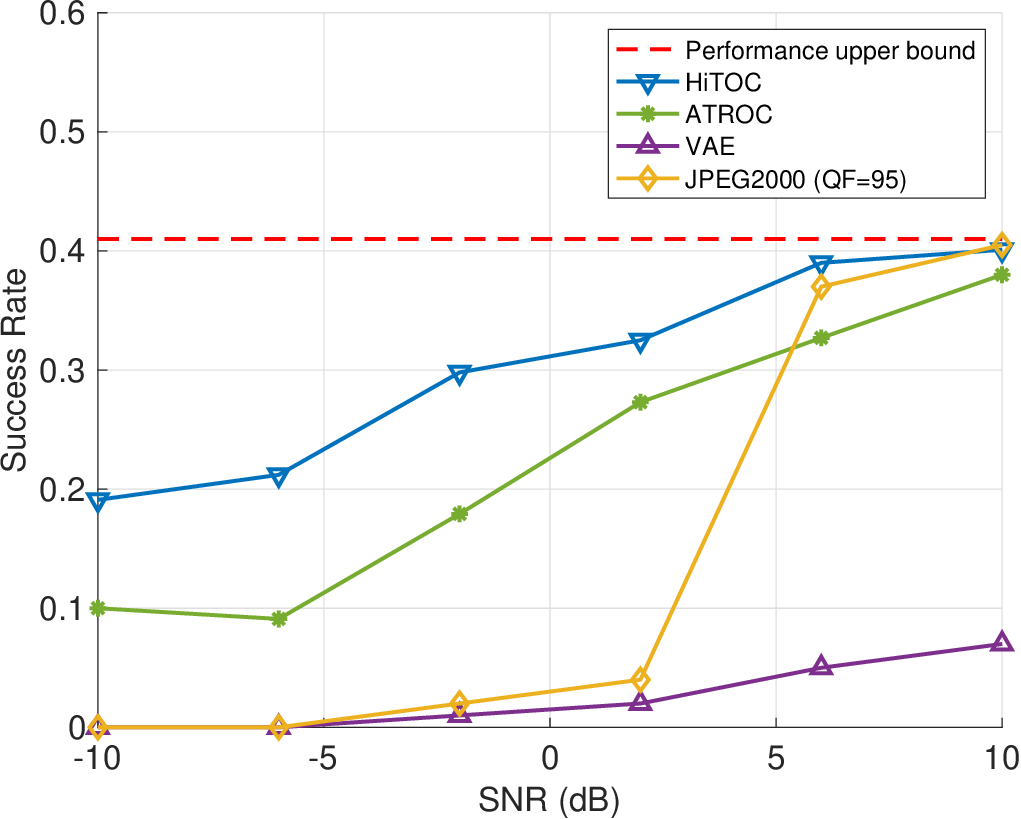}
        \caption{}
        \label{fig:success_rate_Rayleigh}
    \end{subfigure}
    \hfill
    \begin{subfigure}[b]{0.24\textwidth}
        \centering
        \includegraphics[width=\textwidth]{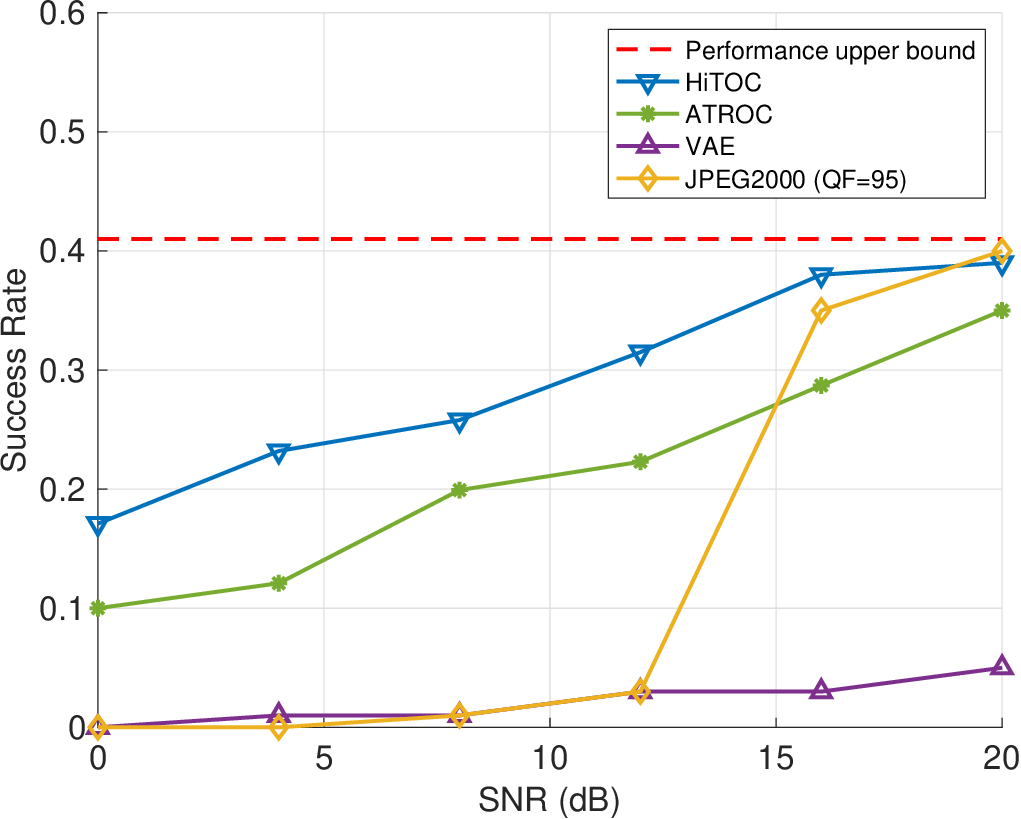}
        \caption{}
        \label{fig:success_rate_AWGN}
    \end{subfigure}
    \vspace{-10pt}
    \caption{Comparison of the success rate versus SNR under (a) AWGN and (b) Rayleigh fading channel models across different baselines.}
    \label{fig:task_performance_SNR}
    \vspace{-10pt}
\end{figure}

The task-specific images under a Rayleigh fading channel with a signal-to-noise ratio (SNR) of 0 dB are shown in Fig.~\ref{fig:visualization} for the subtasks of (a) ``Navigate to the tomato,'' and (b) ``Locate to the fridge.'' In HiTOC, the tomato and the floor are preserved more clearly in Fig.~\ref{fig:visualization}(a) when comparing with the other schemes. In contrast, in Fig.~\ref{fig:visualization}(b), the tomato becomes less visible as the model shifts attention to the fridge. On the other hand, ATROC does not utilize subtask information, so it attempts to preserve all possible visual features that may relate to actions. As a result, the tomato appears less distinct in Fig.~\ref{fig:visualization}(a) but more visible in Fig.~\ref{fig:visualization}(b). The VAE baseline only retains coarse shapes and general colors and the JPEG2000 baseline fails to reconstruct images under such low SNR conditions.
In Fig.~\ref{fig:task_performance_SNR}, we compare the success rate versus SNR across under (a) AWGN and (b) Rayleigh fading channel models for different baselines, respectively. The proposed HiTOC framework preserves subtask-relevant features in the transmitted images. It achieves 5.5\% and 11\% higher success rate than ATROC at an SNR of 10 dB and 20 dB under AWGN and Rayleigh fading channels, respectively. When the SNR is high, traditional coding schemes can almost fully reconstruct the image, resulting in performance comparable to directly transmitting the raw image. As the SNR decreases, JPEG2000 fails to recover the image, causing a significant drop in success rate. Although HiTOC requires subtask information for image encoding and decoding, Table~\ref{tab:bits_service} shows that it increases the number of transmitted bits by only 3\% when compared with ATROC and VAE. Moreover, HiTOC reduces the number of  transmitted bits by 83\% when compared with JPEG2000.
\begin{table}[t]
    \vspace{6pt}
    \centering
    \caption{\small\textsc{Bits per Image for Each Scheme}}
    \label{tab:bits_service}
    \begin{tabular}{|c|c|c|}
        \hline
        \textbf{HiTOC} & \textbf{ATROC / VAE} & \textbf{JPEG2000 (QF = 95)} \\
        \hline
        152{,}064 & 147{,}456 & 884{,}000 \\ 
        \hline
    \end{tabular}
    \vspace{-15pt}
\end{table}

\section{Conclusion}
\label{sec:conclude}
In this paper, we proposed the HiTOC framework, which leveraged LLMs hosted on an edge server to control a robot for long-horizon tasks. The high-level planner LLM decomposed a task into multiple subtasks. We used a conditional module to encode the subtask information. The JSCC encoder and decoder were conditioned on this information to adaptively preserve features relevant to each subtask. The low-level actor LLM then chose an action based on the task-specific image. We further introduced a cVIB approach to ensure that the encoded subtask information contained action-relevant features, while the encoded task data, conditioned on the current subtask, retained the action-related information. Simulations in the AI2-THOR environment showed that the proposed HiTOC framework outperformed three state-of-the-art baselines in terms of success rate for long-horizon tasks. For future work, we plan to extend our framework to multi-agent scenarios.
\vspace{-5pt}

\bibliographystyle{ieeetr}
\bibliography{refs}
\end{document}